\documentclass[preprint]{emulateapj} 
\usepackage{amsmath}
\usepackage{amsfonts}
\usepackage{amssymb}
\usepackage{natbib}
\usepackage{graphicx}

\begin{document} 

\title{A UV-to-MIR monitoring of DR Tau: exploring how water vapor in the planet formation region of the disk is affected by stellar accretion variability} 

\author{A. Banzatti\altaffilmark{1,3}, M. R. Meyer\altaffilmark{1}, C. F. Manara\altaffilmark{2}, K. M. Pontoppidan\altaffilmark{3}, L.Testi\altaffilmark{2}} 
\altaffiltext{1}{ETH Z\"urich, Institut f\"ur Astronomie, Wolfgang-Pauli-Strasse 27, CH-8093 Z\"urich, Switzerland} 
\altaffiltext{2}{European Southern Observatory, Karl Schwarzschild Str. 2, D-85748 Garching bei M\"unchen, Germany} 
\altaffiltext{3}{Space Telescope Science Institute, 3700 San Martin Drive, Baltimore, MD 21218, USA} 
\email{banzatti@stsci.edu}

\begin{abstract} 
Young stars are known to show variability due to non-steady mass accretion rates from their circumstellar disks. Accretion flares can produce strong energetic irradiation and heating that may affect the disk in the planet formation region, close to the central star. During an extreme accretion outburst in the young star EX Lupi, the prototype of EXor variables, remarkable changes in molecular gas emission from $\sim1$ AU in the disk have recently been observed \citep{banz}. Here, we  focus on water vapor and explore how it is affected by variable accretion luminosity in T Tauri stars. We monitored a young highly variable solar-mass star, DR Tau, using simultaneously two high/medium-resolution ESO-VLT spectrographs: VISIR at 12.4 $\mu$m to observe water lines from the disk, and X-shooter covering from 0.3 to 2.5 $\mu$m to constrain the stellar accretion. Three epochs spanning timescales from several days to several weeks were obtained. Accretion luminosity was estimated to change to within a factor $\sim2$, and no change in water emission was detected at a significant level. In comparison to EX Lupi and EXor outbursts, DR Tau suggests that the less long-lived and weaker variability phenomena typical of T Tauri stars may leave water at planet-forming radii in the disk mostly unaffected. We propose that these systems may provide evidence for two processes that act over different timescales: UV photochemistry in the disk atmosphere (faster) and heating of the disk deeper layers (slower). 
\end{abstract} 


\section{INTRODUCTION} \label{sec:intro}
The detection of warm gas emission from the inner few AU of young circumstellar disks \citep{sal08,cn08} has opened a new window on the chemical and physical characterization of the environments where rocky planet properties are probably set. Surveys of mid-infrared (MIR) molecular emission from $\sim$100 circumstellar disks over a range of spectral types and evolutionary stages have already moved important steps forward in this context \citep{pont10a,cn11,sal11}. The observed emission from water, OH, and organic molecules (HCN, C$_{2}$H$_{2}$, and CO$_{2}$) comes from warm gas (300-1000 K) in the planet-forming zone of the disk, and is generally found to be common in T Tauri disks, while reduced or absent in more massive (Herbig Ae/Be) or more evolved (transitional) disks. Moreover, molecular abundances are found to be not simply inherited from the prestellar phase, but show evidence for an ongoing active chemistry. The infrared molecular gas emission is a privileged tracer of the processing of warm material in evolving protoplanetary disks. Studies of the interplay between accretion histories and the molecular chemistry are essential to our understanding of disk evolution and planet formation, and could perhaps provide hints to help explain the diversity in composition of (exo)planets.

An often overlooked element in disk evolution is the effect of non-steady accretion processes during star formation. In this regard, an exceptional experiment was provided by a recent extreme accretion outburst in EX Lupi \citep[the prototype of EXor variables,][]{herb89,herb08}. This is a classical T Tauri star showing eruptive accretion phenomena that in strength are in between FUor outbursts and the typical variability found in the T Tauri phase of young stars. From comparison of MIR spectra taken during outburst and in a preceding quiescent phase, \citet{banz} found that while the molecular spectrum in quiescence showed water, OH, and organic emission typical of T Tauri systems, it changed remarkably in outburst. Water emission increased, suggesting a larger extent of the warm emitting gas caused by the increased disk heating. On the other hand, detection of previously unseen OH lines suggested ongoing OH production via UV photodissociation of water. Strikingly, emission from all organics disappeared. EX Lupi provided the first evidence that the warm molecular gas emission can change dramatically, in the \textit{same} disk, depending on the accretion phase. If the EXor and classical T Tauri variability are phases of the same accretion history that all forming stars undergo \citep{hartm}, effects similar to those observed in EX Lupi might be common in T Tauri stars. To better understand both the properties and the evolution of the molecular gas in the planet formation region of \textit{all} circumstellar disks we may need to consider the time domain, especially during the non-steadily accreting phases of the star-disk interaction. This work was initiated to probe this idea.

In this paper we study the emission observed toward the active T Tauri star DR Tau. Situated in the Taurus-Auriga star-forming region at 140 pc from Earth \citep{taurus}, DR Tau is one of the most studied classical T Tauri stars. Yet, it is one of the most peculiar as well. It showed a slow brightness increase over $\sim20$ years between 1960 and 1980 \citep{chav,gotz}, which raised it from being a relatively faint star ($\lesssim14$ mag in $V$ band) to one of the brightest stars in the Taurus association ($\sim11$ mag in $V$ band). Its behavior was attributed to strong non-steady accretion \citep{bert88}, and DR Tau was included in the first list of EXor variables by \citet{herb89} together with EX Lupi. Since 1980, DR Tau has mantained its enhanced brightness while showing large photometric and spectroscopic variability \citep[e.g.,][]{alencar,grankin}. The high and variable veiling produced by accretion has made its spectral classification challenging and dubious. However, recent studies performed in lower veiling phases agree on a K5-K7 spectral type \citep{mora,petrov}. DR Tau is still a young system ($\sim1$--2 Myr) with a relatively massive disk of $\sim0.01M_{\odot}$ \citep{ricci10a}. It is one of the first disks where warm water vapor emission was detected \citep{sal08}. Together with water, OH and organic molecules have been detected at both mid- and near-infrared (NIR) wavelengths \citep{pont10a,mand,brown13}. Given its rich molecular emission and its strong and variable accretion, it is an optimal target for investigating how the former is affected by the latter. In this work, we monitored DR Tau through its MIR water emission lines from the inner disk using the ESO Very Large Telescope (VLT) \textit{Imager and Spectrometer for the mid-InfraRed} \citep[VISIR,][]{visir}. Simultaneously, the stellar accretion emission was monitored using VLT/X-shooter \citep{xshooter}.

\begin{deluxetable}{l c c c c c}
\tabletypesize{\small}
\tablewidth{0pt}
\tablecaption{\label{tab:obs}Three epochs of VISIR observations of DR Tau} 
\tablehead{ \colhead{Epoch}  & \colhead{Airmass} & \colhead{Seeing} & \colhead{Doppler} & \colhead{Frames} & \colhead{t$_{int}$} \\ &  & (arcsec) & (km/s) & & (s) }
\startdata
 \textbf{1:} 27 Nov 2011 & 1.38 & 0.4--0.5 & 21.15 & 32 & 2240 \\
 \textbf{2:} 02 Dec 2011  & 1.51 & 0.4--0.7 & 23.87 & 40 & 2800 \\
 \textbf{3:} 14 Jan 2012  & 1.41 & 0.4--0.5 & 44.64 & 41 & 2870 
\enddata
\tablecomments{Details on the values reported in this table are given in Section \ref{sec: obs_visir}. Airmasses are averaged per epoch. The seeing is estimated at 12.4 $\mu$m using the measured width of the source PSF. Doppler shifts are measured with respect to rest frequencies of the targeted emission lines. The number of frames and the respective on-source integration time of the reduced spectra are given in the last two columns.}
\end{deluxetable} 

\begin{figure}
\includegraphics[width=0.5\textwidth]{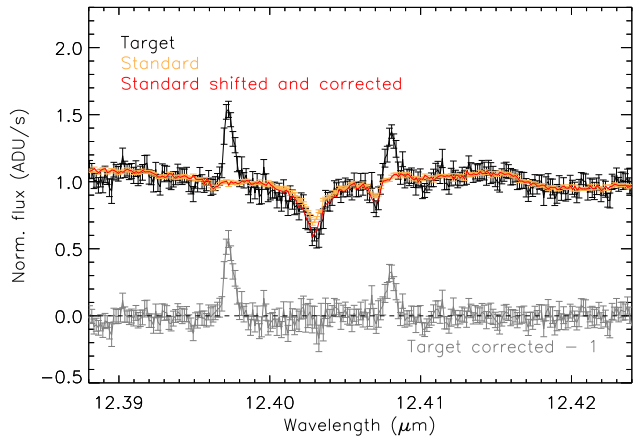} 
\caption{Telluric correction of VISIR spectra. For illustration, we show here the second epoch of observations obtained in this program. DR Tau is shown in black/grey (before/after telluric correction), HD50138 in orange/red (before/after cross-correlation with DR Tau and airmass correction). The two relevant telluric lines used for cross-correlation as explained in the text are at 12.403 and 12.407 $\mu$m. The 12.396-$\mu$m water line has hardly any telluric counterpart, due to its high upper level energy ($\sim5800$ K).}
\label{fig:tell_corr}
\end{figure}

\section{OBSERVATIONS} \label{sec: obs}
This monitoring program (088.C-0666, PI: Banzatti) was carried out in service mode using two instruments at the VLT: VISIR (at VLT/UT3) and X-shooter (at VLT/UT2). As part of this program, DR Tau was observed on six nights between November 2011 and January 2012. In three out of these six nights, simultaneous observations with the two VLT instruments were performed, and these are the three epochs studied in this paper\footnote{The X-shooter epochs, six in total, will be analyzed together in a future paper.}  (see Table \ref{tab:obs}). The first and second epochs were taken 5 days apart; the third was taken 43 days after the second. Such a scheduling spanning different time ranges was proposed in an attempt of observing the star in different phases of its unpredictably variable accretion. However, the exact choice of dates depended on atmospheric conditions, which we required to be photometric\footnote{We required clear sky, seeing better than 0.8\arcsec, and precipitable water vapor column of less than 2 mm.}, and the challenging scheduling of the two VLT instruments, which we required to be used simultaneously. In each epoch, DR Tau was observed for $\sim2$ hours with VISIR, in good atmospheric and airmass conditions (see Table \ref{tab:obs}). Given that DR Tau is a bright star, the X-shooter observations required a much shorter time ($\sim15$ min) and were performed during the first quarter of each VISIR epoch. 

Observing water emission lines from Earth is difficult because every targeted line has its telluric counterpart. The observations were done in November-January to maximize the Doppler shift between the target and Earth, such that the targeted lines were shifted by $\sim20$--45 km/s from their telluric counterparts (Table \ref{tab:obs}). We chose to observe high-energy lines (at 12.396 and $12.407 ~ \mu$m, upper level energy $E_u$ 5000-5800 K), to ensure weak telluric counterparts and minimize absorption effects even in the case of partial overlap of line wings (see Figure \ref{fig:tell_corr}). The line sample was restricted by a limited availability of filters along with a need for long integrations to reach sufficient signal-to-noise ratios (S/N), to be able to probe variations in emission from one epoch to the other. Given these constraints, one VISIR setting per epoch was observed, centered at 12.407 $\mu$m.

\begin{figure*}
\includegraphics[width=1\textwidth]{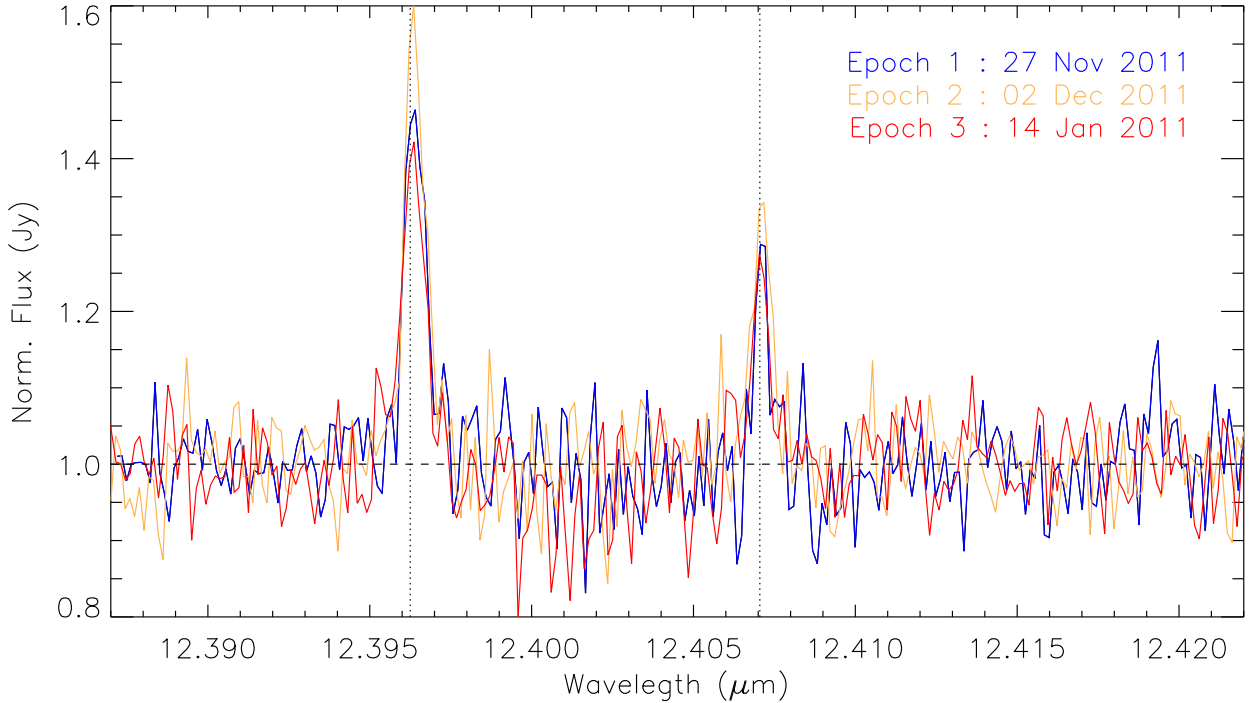} 
\caption{Three epochs of VISIR data, with the two targeted water lines (rest frequencies are marked with dotted lines). The spectrum in each epoch was reduced and telluric-corrected as described in Section \ref{sec: obs}. Corrections for the heliocetric-baryocentric velocity of the VLT at Paranal and the radial velocity of the star \citep[23 km/s,][]{app88,petrov} has been applied to each epoch. The spectra are continuum-normalized to illustrate the relative change in water-to-continuum emission from epoch to epoch. }
\label{fig:visir}
\end{figure*}

\subsection{VISIR spectra} \label{sec: obs_visir}
VISIR was used in the cross-dispersed mode, providing a resolving power R $\sim20,000$ ($\Delta v \sim$15 km/s) at 12.4 $\mu$m. The data were taken using standard chopping and nodding techniques for VISIR\footnote{A detailed description of observing techniques is available in the VISIR User Manual online on www.eso.org.}. The telescope nods parallel to the slit between positions on-source (A) and off-source (B) with a sequence ABBA, and a secondary mirror chops parallel to the slit with a sequence ABAB starting in the nod position A, and BABA starting in the nod position B. Each VISIR raw datacube provided by ESO includes the four half-chop-cycle images performed in one nod position (e.g., A$_{i}$, B$_{i}$, A$_{i+1}$, B$_{i+1}$, where $i$ is the chop cycle number). The slit used in the high resolution mode (width = $0.75\arcsec$, length = $4.1\arcsec$) does not allow chopping on-slit, as the minimum chopping throw is $8\arcsec$. Therefore, the B positions do not have the source on the slit and are used only for background subtraction. This effectively reduces the on-source integration time of VISIR in the cross-dispersed mode to half the total time, bringing it to 2940 s in each of our epochs. 

The VISIR data were reduced using a suite of self-made IDL routines intended to optimize the standard VISIR data reduction of the ESO Recipe Execution Tool (EsoRex), version 3.9.0. First, small shifts (typically fractions of a pixel) caused by mechanical oscillations of the grating between sequential images needed to be corrected. To do that, in each nod position the individual A and B images were combined, applying the shift estimated by cross-correlation of two telluric lines observed in the region between the targeted water lines. In our data, this procedure worked well in reducing artifacts from sky residuals that would otherwise affect the targeted water emission, if processed only through the standard ESO pipeline \citep[a similar correction was implemented by][]{carmona}. 
Standard EsoRex procedures were then used to perform image combination, distortion correction, and wavelength calibration as follows. Sequential nod pairs were combined into nod half-cycle images by averaging A and $-$B. In total, each epoch has 42 such AB frames, whose on-source integration time is 70$~$s each. We carefully monitored the observing conditions frame by frame, by means of measuring the peak and the width of the gaussian point-spread-function (PSF) fitted to the source trace on the detector, in the spatial direction. An estimate of the atmospheric seeing at 12.4 $\mu$m ($N$-band) was based on the measured width of the PSF, and is reported in Table \ref{tab:obs}. Variable atmospheric conditions can overwhelm the target signal in a given frame, when the sky infrared background varies too much from A to B. The actual number of frames that could be used in each epoch is therefore smaller than the total and differed from epoch to epoch (see Table \ref{tab:obs}). 
Nod images were then corrected through pixel interpolation for small distortions caused by the instrument, using the EsoRex parameters \texttt{slit\_skew} = $1^{\circ}$ and \texttt{spectrum\_skew} = $0.4^{\circ}$, which were found to be appropriate in the 12.4 $\mu$m setting. Wavelength calibration of individual spectra was done by cross-correlation with an atmospheric model set on typical Paranal conditions, using the same telluric absorption lines mentioned above. 

An additional background subtraction was needed to correct for sky background residuals produced by variable atmospheric water vapor within chop cycles. This was performed by a median subtraction row by row in the cross-dispersed direction, as in \citet{pont10b}. An individual spectrum was then extracted from each combined (AB) nod frame using optimal extraction to maximize S/N \citep{horne}. The individual spectra in each epoch were then combined into a weighted average yielding one spectrum per epoch. Flux errors on individual pixels in the averaged spectrum were set to the standard deviation of the weighted mean of pixel values as measured in the total number of frames in each epoch. 

Telluric absorption had a negligible influence on the two targeted water lines during this monitoring of DR Tau, for the reasons explained at the beginning of this section. Nonetheless, a telluric standard was observed and utilized to correct also for the inhomogeneous detector response (as flat fields are usually not taken for VISIR). We used a Herbig Ae/Be star (HD50138), which was observed after DR Tau each night, selected on the basis of its brightness and absence of emission lines at 12.4 $\mu$m. The spectra were normalized to their continua, aligned by cross-correlation, and then divided (DRTau/HD50138), after airmass ($A$) correction of HD50138 with respect to DR Tau. To do that, an exponential correction was applied to the telluric absorption lines in the HD50138 spectrum, as $f_{corr} = f_{obs}^{(A_{DRTau}/A_{HD50138})}$. The match between HD50138 and DR Tau after this correction was found to be excellent in all epochs. Figure \ref{fig:tell_corr} illustrates the telluric correction procedure, while Figure \ref{fig:visir} shows the three epochs of the final VISIR spectra, reduced and corrected as described in this section.

\begin{figure}
\includegraphics[width=0.5\textwidth]{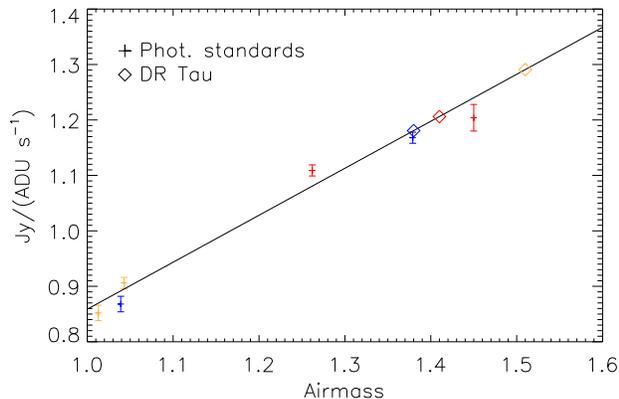} 
\caption{Relation between airmass and conversion factor Jy/(ADU s$^{-1}$) as derived from photometric standard stars (observed before and after DR Tau in each epoch). Conversion factors as measured from individual standards are shown with crosses (plus error-bars), while the black line is a linear fit to them. The values extrapolated for DR Tau, using the average airmass in each epoch, are shown with diamonds. Epochs are color-coded as in Figure \ref{fig:visir}.}
\label{fig:convfact}
\end{figure}

To perform absolute flux calibration of the VISIR epochs, we used photometric standard stars that were observed before and after DR Tau each night. HD22663, HD6805, HD7055, and HD37160 were observed as part of the standard ESO calibrators, and in addition we observed Sirius A. From knowledge of their flux in Jy, a conversion factor Jy/(ADU s$^{-1}$) can be derived using the measured detector counts for each standard, which were observed at different airmasses. Figure \ref{fig:convfact} shows a linear fit to the conversion factors estimated over the three epochs using two standards each night. The model fits the individual epochs reasonably well, supporting the idea of overall stable atmospheric conditions. The continuum in DR Tau was then estimated using the fitted relation between airmass and conversion factor, extrapolated at the average airmass of DR Tau in each epoch (Figure \ref{fig:convfact}). The estimated continuum decreases by 25\% in the second epoch, and rises by 23\% in the third. In the second epoch, the seeing was relatively worse (but still lower than the slit width, ensuring limited slit losses; see Table \ref{tab:obs}), and the two photometric standards were taken at $\sim0.4$--0.5 lower airmasses with respect to DR Tau. To check whether the decrease in continuum level could be simply due to the general worse seeing conditions, we used the technique described above to estimate the continuum also for the telluric standard HD50138. We found the same trend, a decrease in the second epoch followed by an increase in the third, but confined to within 5\%, comparable to the precision obtained on the photometric standards. We therefore concluded that variable atmospheric conditions may have affected by at most 5\% the continuum in DR Tau, but that it intrinsically changed by $\sim20\%$. This is consistent with the change estimated in the $H$ band from the X-shooter data (see Section \ref{sec:disc}).

\begin{figure*}
\includegraphics[width=1\textwidth]{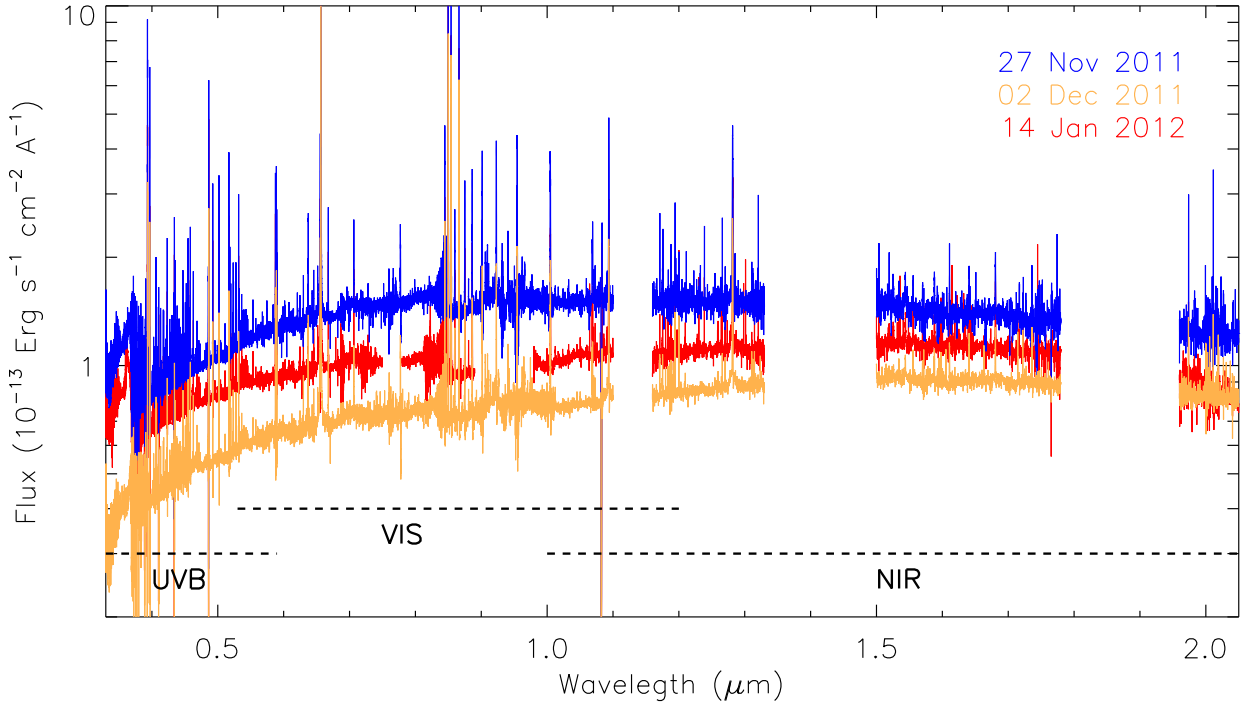} 
\caption{Three epochs of X-shooter spectra, with the three arms covering from UVB to NIR. Each spectrum was taken simultaneously to the VISIR spectrum of a given epoch (Figure \ref{fig:visir}). The data were reduced and calibrated as explained in Section \ref{sec: obs_xsh}.}
\label{fig:xshooter}
\end{figure*}

\subsection{X-shooter spectra} \label{sec: obs_xsh}
The X-shooter echelle spectrograph observes in three spectral arms, providing a simultaneous combined coverage of $\sim$300--2500 nm: a UVB arm covering $\sim$300--590  nm, a VIS arm covering $\sim$530--1020 nm, and a NIR arm covering $\sim$1000--2480 nm. In each epoch, we asked for two settings: the wide-slit mode (5\arcsec) with 20 s exposure time, and the narrow-slit mode ($0.4\arcsec$ in the VIS and NIR arms, $0.5\arcsec$ in the UVB arm) with 120 s exposure time. Four nodding repetitions with the ABBA scheme were done to obtain an optimal background subtraction. The narrow-slit setting provided the higher-resolution spectra (R = 9100, 17400, 10500 in the UVB, VIS, and NIR arms, respectively) used to monitor accretion emission in this work. The wide-slit setting avoids flux losses and was taken to perform the spectro-photometric calibration of the higher-resolution spectrum in each epoch. The data were reduced using EsoRex and the ESO pipeline for X-shooter, version 1.3.7 \citep{xshpip}. The pipeline applies standard reduction procedures to perform background subtraction, spectral extraction, wavelength calibration, and flux calibration in each spectral arm separately. A more accurate absolute flux calibration of the higher-resolution spectra was then obtained by scaling each spectral arm to the flux level measured using the wide-slit mode. The overlapping spectral regions between the arms were used to check the results, which were found to agree well in all arms. Finally, the spectroscopic standard stars provided by ESO were used to perform telluric correction through the IRAF task \textit{telluric}. Figure \ref{fig:xshooter} shows the three epochs of the higher-resolution X-shooter spectra, reduced and calibrated.

\begin{deluxetable*}{c c c c c c c}
\tabletypesize{\small}
\tablewidth{0pt}
\tablecaption{\label{tab:lines} Properties of water lines observed with VISIR.}
\tablehead{\colhead{Wavelength} & Transition & $A_{ul}$ & $E_{u}$ & &  \colhead{Line flux} \\ \colhead{ ($\mu$m)} & $J_{\:K_a\:K_c}$ &  (s$^{-1}$) &  (K) & Epoch 1 & Epoch 2 & Epoch 3}
\tablecolumns{6}
\startdata
12.39625 & $17_{\:4\:13} \rightarrow 16_{\:3\:14}$ & 7.7 &  5781 &  1.80 $\pm$ 0.19 & 1.60 $\pm$ 0.18 & 1.46 $\pm$ 0.16 \\
12.40708 & $16_{\:3\:13} \rightarrow 15_{\:2\:14}$ & 4.2 &  4945 &  0.91 $\pm$ 0.17 & 0.92 $\pm$ 0.12 & 0.81 $\pm$ 0.13 
\enddata
\tablecomments{Line fluxes are given in $10^{-14}$ erg s$^{-1}$ cm$^{-2}$. Line properties are taken from the HITRAN database.}
\end{deluxetable*} 

\section{ANALYSIS} \label{sec:res}
\subsection{Water line properties} \label{sec:res_wat}
The two targeted water lines are both well detected in all epochs (see Table \ref{tab:lines}), and present a single-peak unresolved profile (Figure \ref{fig:visir}). This is consistent with gas emission from the inner regions of a nearly face-on disk (inclination $\sim10^{\circ}$, where $0^{\circ}$ is face on), as proposed for DR Tau by \cite{pont11} and \cite{brown13}  from spectro-astrometry of CO and NIR water emission lines respectively. Assuming Keplerian rotation and a stellar mass of 0.8 M$_{\odot}$ \citep{ricci10a}, a lower limit to the emitting radius of the 12.4 $\mu$m water vapor in DR Tau is found at $\gtrsim 0.1$ AU from our VISIR data. We measured the Doppler shift between the observed line peaks and their rest frequencies \citep[taken from the HITRAN database,][]{hitran} after correction for the known heliocentric-baryocentric velocity in each epoch provided by ESO for the VLT at Paranal. The heliocentric velocity of both water lines was found to be $\sim25$ km/s in all epochs of our VISIR data. This is consistent with the radial velocity of the star as estimated in 1987 by \citet{app88} and in 2007--2010 by \citet{petrov}, who measured an average of $23\pm2$ km/s using stellar photospheric lines. 

We fit the observed water emission using gaussian functions on top of a linear continuum, estimated over the spectral range covered by the VISIR setting and excluding the region of the targeted emission lines and of telluric absorption lines (12.393--12.409 $\mu$m). For line fitting, we use the least-squares fitting routines by \citet{mark}. We estimate each line flux and FWHM from the peak and the width of the best-fit gaussian function, and the flux error is propagated from the uncertainty on the best-fit parameters. Fitted line widths are consistent with the nominal VISIR resolution of 15 km/s within the errors, and we therefore fix this value in all epochs. Line fluxes are measured in the normalized spectrum of each epoch, and then scaled to the continuum level at 12.4 $\mu$m as estimated in Section \ref{sec: obs_visir}. The line-to-continuum ratio, or the line flux in the normalized spectrum, increases by $\sim30\%$ in the second epoch and decreases by roughly the same amount in the third. This contrast ``flickering" is observed in both water lines at roughly the same level (Figure \ref{fig:visir}). We therefore estimate the average line flux change between epoch $i$ and epoch $i+1$ as
\begin{equation}
\left( \dfrac{line_{a,i+1}}{line_{a,i}} + \dfrac{line_{b,i+1}}{line_{b,i}} \right) /2 \, ,
\end{equation}
where the two water lines are labelled as $a$ and $b$. Calibrated line fluxes and their propagated errors are reported in Table \ref{tab:lines}, while their relative changes between epochs are shown in Figure \ref{fig:variations}.

\subsection{Accretion luminosity} \label{sec:res_acc}
We derive the accretion luminosity ($L_{acc}$) from estimates of the UV excess emission in the Balmer jump region in X-shooter spectra, following previous work \citep{valenti,herczeg,elisab}. The model and method we adopt are presented in detail in \citet{Manara13b}; here we give a basic description. We fit the data using a model with three components: a photospheric template, a reddening law, and a model for the accretion spectrum. As the photospheric template we use a Class III young stellar object (2MASS J05390540-0232303) with spectral type K7, similar to DR Tau, from the suite of photospheric templates collected in \citet{carlo}. We adopt the reddening law from \citet{card} with $R_V=3.1$ \citep[appropriate for Taurus,][]{herczeg} and explore values of $A_V$ within 0 and 3 mag, in steps of 0.1 mag. 
We model the continuum emission of the accretion spectrum as an isothermal slab of atomic hydrogen gas, which is an approximation adequate for young accreting stars \citep[e.g.,][]{valenti,herczeg}. It includes bound-free and free-free emission from both H and H$^{-}$, assumes local thermodynamic equilibrium (LTE) conditions, and is described by three parameters: the electron temperature (T$_{\rm{slab}}$), the electron density ($n_e$), and the optical depth at 300 nm ($\tau_{300}$), which is related to
the length of the slab. T$_{\rm{slab}}$ is explored in the range 5000--11000 K, $n_e$ in the range $10^{11}$--$10^{16}$ cm$^{-3}$, and $\tau_{300}$ in the range 0.01--5, which are typical for this kind of description. Additional parameters of the model are two normalization constants: $K_{\rm{phot}}$ accounts for the differences in distance and radius between DR Tau \citep[for the stellar luminosity we adopt $L_{\star}=0.85 ~ L_{\odot}$ from][]{james03} and the photospheric template star, and $K_{\rm{slab}}$ accounts for the area of the accretion slab at the stellar surface, as if $L_{acc}$ was produced by a hot spot. In total, the free model parameters are five: $A_V$, T$_{\rm{slab}}$, $n_e$, $\tau_{300}$, and $K_{\rm{slab}}$. 

We fit the data in several spectral ranges in the Balmer and Paschen region (330--480 nm), including the Balmer jump, the slopes of the Balmer and Paschen continua, and the continuum at $\sim$710 nm. The best fit is found by minimization of a $\chi^2$-like function defined as the sum of the squared deviations (data -- model) divided by the error. The goodness of the best-fit accretion model is checked also visually, by looking at the degree of veiling produced in some photospheric absorption features (namely CaI at 420 and 616 nm, and the TiO lines at 844 nm). More details on this model and the fit methodology can be found in \citet{Manara13b}.

\begin{figure}
\includegraphics[width=0.5\textwidth]{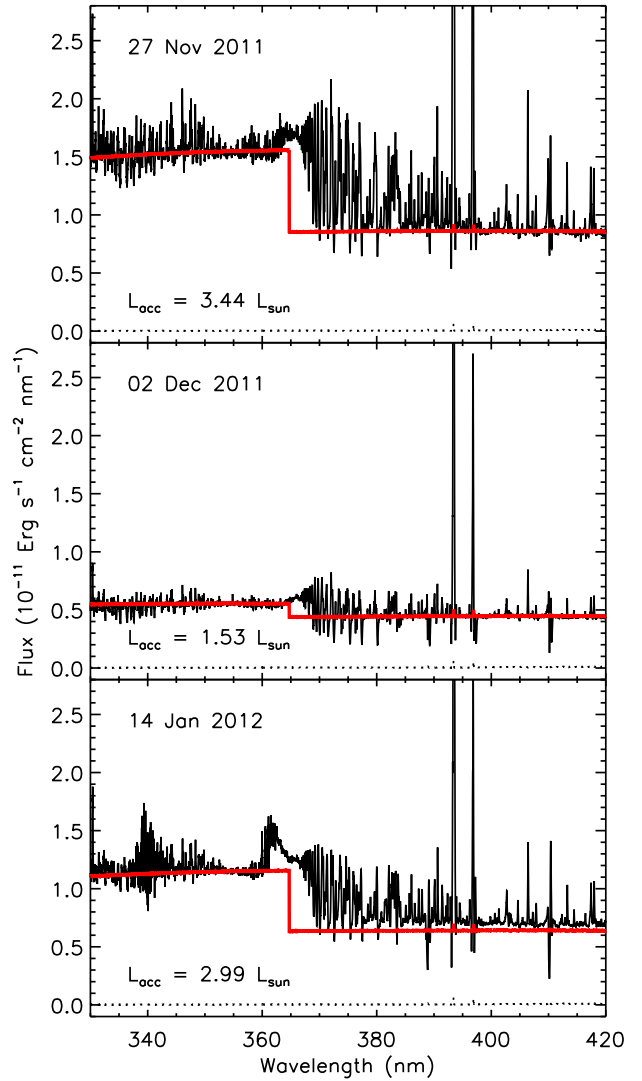} 
\caption{Best fit for the accretion model in each epoch, derived as explained in Section \ref{sec:res_acc}. In each plot, the X-shooter spectrum of DR Tau is shown in black, the photospheric template in dotted line, and the fitted model (slab + photosphere) in red. The photospheric contribution is three orders of magnitude smaller than the accretion slab contribution in all epochs.}
\label{fig:BJfit}
\end{figure}

The extinction $A_V$ is found to be 1.8--1.9 mag in this monitoring of DR Tau and is consistent with being the same in all three epochs. This can be understood in terms of the almost face-on disk geometry (see Section \ref{sec:res_wat}), where changes in accretion columns onto the star would be unlikely to produce large changes in extinction \citep[see also][]{alencar}. We therefore decided to fix $A_V$ to a value of 1.8 mag common to all epochs. This simplification is convenient because the largest uncertainty on the derived $L_{acc}$ is due to the uncertainty in $A_V$, but in a relative sense (comparing epoch to epoch), the uncertainty on the change in $L_{acc}$ is smaller. Thus we reduce the free parameters to T$_{\rm{slab}}$, $n_e$, $\tau$, $K_{\rm{slab}}$ and evaluate the change in the best fit from epoch to epoch as if it were primarily given by accretion. After finding a best-fit slab model in each epoch, $L_{acc}$ is derived as $L_{acc} = 4\pi d^2 F_{acc}$, where $d$ is the distance to the source and $F_{acc}$ is the flux integrated over the 50--2500 nm range in the slab model rescaled with $K_{\rm{slab}}$. Figure \ref{fig:BJfit} shows the best-fit models obtained in the three epochs of X-shooter data. The uncertainty on $L_{acc}$, as derived from the uncertainty on the slab model parameters, is of the order of 10--20\%. 

\section{DISCUSSION} \label{sec:disc}
\begin{figure}
\includegraphics[width=0.5\textwidth]{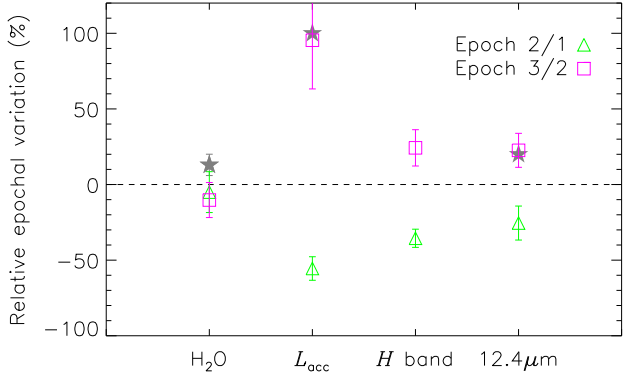} 
\caption{Relative variations of different tracers (12.4-$\mu$m water line fluxes, accretion luminosity, continuum level in the $H$ band and at 12.4 $\mu$m), estimated in this monitoring study of DR Tau (see Section \ref{sec:res}). We compare epoch 2 to 1 and epoch 3 to 2 in a time sequence, taking the ratio of their tracers. Grey stars indicate the changes measured in EX Lupi \citep{asp,banz} divided by 20 (see Section \ref{sec:disc}).}
\label{fig:variations}
\end{figure}

Figure \ref{fig:variations} summarizes our results on the relative changes in water emission and stellar accretion, estimated in this monitoring study of DR Tau as described in Section \ref{sec:res}. The accretion luminosity decreased by $\sim55\%$ in the second epoch, and increased by $\sim95\%$ in the third. The continuum in the X-shooter spectra follows this trend (see Figure \ref{fig:xshooter}), supporting the idea that its variation was related to the change in accretion. For illustration, we show in Figure \ref{fig:variations} the changes in the $H$-band continuum level, measured as the average pixel ratio between epoch $i+1$ and epoch $i$ in the range 1.52--1.72 $\mu$m. Using the absolute continuum estimated from the VISIR data (Section \ref{sec: obs_visir}), the 12.4\,-$\mu$m flux shows the same trend, decreasing in the second epoch and increasing in the third within $\sim$20--25\% (consistent with the variation in the $H$-band continuum). The change of $\sim30\%$ in the continuum-normalized water line fluxes (Figure \ref{fig:visir}) can therefore be attributed to the change, similar in fraction but opposite in sign, of the $N$-band continuum level. In other words, the calibrated water line fluxes are consistent in the three epochs within the errors. In conclusion, while in this monitoring program of DR Tau the accretion luminosity changed within a factor $\sim2$, no change in water emission was detected at a significant level (a 10\% decrease in line fluxes is seen by comparison of epoch 3 to 2 but only at a 1$\sigma$ level, see Figure \ref{fig:variations}).

\begin{figure}
\includegraphics[width=0.5\textwidth]{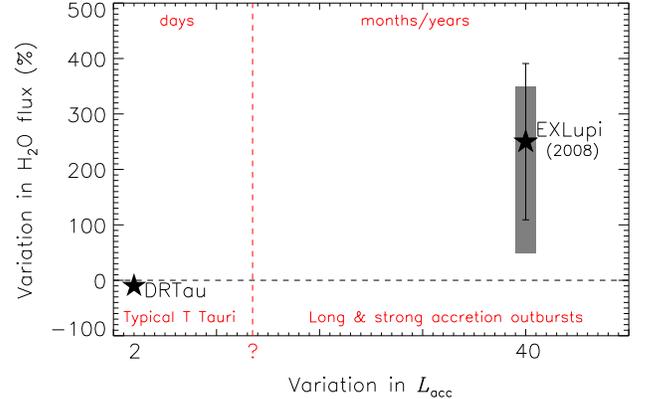} 
\caption{Comparison of DR Tau (this work) and EX Lupi \citep{banz} in terms of changes in accretion luminosity and water vapor emission at 12.4 $\mu$m (black stars). The grey-shaded area shows, for reference, the range in water line flux changes measured by \citet{banz} over the entire \textit{Spitzer} coverage (10--35 $\mu$m). While the variation in $L_{acc}$ observed in DR Tau is typical during the T Tauri phase, and no change in water emission was significantly detected, EX Lupi was studied during a recent extreme accretion outburst, and water emission increased. The breaking point between the two regimes may depend both on the duration and the strength of the accretion variation.}
\label{fig:drtau_exlup_comp}
\end{figure}

Variable water emission from the disk was previously observed from comparison of a quiescent phase to an accretion outburst in EX Lupi \citep{banz}, using spectra obtained with \textit{Spitzer}-IRS \citep{spitzer,irs}. In outburst, \citet{asp} estimated an increase in $L_{acc}$ of a factor $\sim40$, while the flux continuum at 12.4 $\mu$m increased by a factor $\sim5$ and water line fluxes in the range 10--35 $\mu$m increased by $\sim 50$--350\% \citep{banz}. Specifically, the two water lines at 12.396 and 12.407 $\mu$m increased in flux by $\sim$ 250\% \footnote{Line fluxes and errors are estimated by fitting the 12.4 $\mu$m water emission, which is blended in \textit{Spitzer} spectra, with a gaussian for each individual line. The basic method is explained in \citet{banz}, while here we additionally fix the line width to FWHM=$\lambda/750$ \citep[][Banzatti et al. submitted]{naji}.} (see Figure \ref{fig:drtau_exlup_comp}). 
In a thought experiment, if what we observed in DR Tau is a scaled-down version of the effect observed in EX Lupi, for an increase in $L_{acc}$ of a factor $\sim2$ we should have seen an increase in the 12.4 $\mu$m continuum of $\sim20\%$ and an increase in water line fluxes of $\sim$ 13\% (dividing all values by 20). The results presented in this paper for DR Tau are consistent with this scaled-down prediction from EX Lupi (see Figure \ref{fig:variations}). The change in water emission is consistent only within 2$\sigma$ and might show an opposite trend, but the small changes are comparable to the precision reached on the line fluxes and are therefore hard to distinguish from the noise.

The behavior observed in EX Lupi and DR Tau may be understood in terms of two processes and their different timescales: UV photochemistry and disk heating. \citet{bb09} showed that in a typical T Tauri disk model the disk surface is dominated by photodissociation and fast chemical timescales ($\lesssim 10^2$ s), while the deeper cold midplane where most of the mass resides in icy solids reaches chemical equilibrium on timescales of $> 10^4$ s. Studying how protostellar envelopes react to accretion outbursts, \citet{johnst} recently showed that thermal equilibrium between gas and dust through collisional heating requires a few weeks in regions with density comparable to the inner disk regions where the 12.4 $\mu$m water emission likely comes from \citep[$\sim10^{9}$ cm$^{-3}$, e.g.,][]{banz}. In EX Lupi the accretion outburst lasted for $\sim$7 months, long enough to change the thermal disk structure by heating up the disk and shift outward the T $\gtrsim$ 170 K region (the snowline) from $\lesssim0.6$ to $\lesssim1.2$ AU \citep{abra09}, consistent with the change in emitting area of the warm gas estimated by \citet{banz}. Stronger water lines for higher accretion were observed probably due to new water vapor, evaporated from icy solids beyond the snowline \citep[e.g.,][]{ciesla} and/or formed in situ in the gas phase from molecular hydrogen and oxygen \citep{glas,bb09}. The weaker and shorter accretion variability observed in DR Tau (and typical of T Tauri stars) might, instead, have the time to affect only the disk surface layers and its main effect would probably be restricted to a variable UV photochemistry. In this case, weaker water lines might be observed in case of photodissociation. This new monitoring study of DR Tau suggests that during the T Tauri phase, a change in $L_{acc}$ within a factor $\sim2$ over a few days only is probably not enough to affect significantly the water vapor in the inner disk, at least as probed in the $N$ band. Perhaps the strength and duration necessary to drive stronger water vapor emission (and possibly production) as observed in EX Lupi are provided only during the extreme EXor outbursts (Figure \ref{fig:drtau_exlup_comp}).

\section{SUMMARY AND CONCLUSIONS} \label{sec: end}
We investigated the effects of stellar accretion variability on the water vapor at planet-forming radii in circumstellar disks during the T Tauri phase, by means of a new high-resolution UV--to--MIR monitoring of DR Tau. Three epochs of simultaneous VLT/VISIR and VLT/X-shooter spectra were obtained between November 2011 and January 2012. Accretion luminosity was derived from the X-shooter data and found to change within a factor $\sim2$ (decreasing in the second epoch, increasing in the third). No change in water line fluxes at 12.4 $\mu$m was measured at a significant level from the VISIR spectra. The enhancement in water emission observed during an extreme accretion outburst in EX Lupi suggests that accretion can drive water vapor production, possibly in connection with a recession of the snowline \citep{banz}. While the change in accretion between the epochs in DR Tau was not large enough to distinguish changes in water emission from noise, our monitoring suggests that if the change in $L_{acc}$ is confined to within a factor $\sim2$, then water vapor emission as probed at 12.4 $\mu$m may remain mostly unaffected (within $\sim10\%$). 
It would be interesting to explore until when this is still the case: what conditions are mild and stable enough to leave the inner disk undisturbed, and where the breaking point is (Figure \ref{fig:drtau_exlup_comp}). This exploration would require constraining two parameters: the strength and the duration of the accretion outburst. While an increase in water line fluxes of $\gtrsim$ 20-30\% is needed to clearly stand out from the noise with the current instrumentation (and EX Lupi suggests that this might happen at an increase in $L_{acc}$ of a factor $\gtrsim4$), an accretion event duration of at least a few weeks might be necessary to allow for thermal equilibrium at the gas densities probed by the observed water vapor \citep{johnst}. Much is still to be understood in the production/destruction of water vapor as driven by accretion variability. Future observations promise to shed more light on the role of accretion histories in shaping the molecular environments of the planet-formation region in protoplanetary disks.
 \\

AB is grateful to several colleagues for valuable conversations about the results from this work, many of which happened during the PPVI conference in Heidelberg: Ted Bergin, Simon Bruderer, James Muzerolle, and Ruud Visser. The authors acknowledge an anonymous referee for comments that helped improving parts of this manuscript. This work is based on observations made with ESO telescopes at the Paranal Observatory under programme ID 088.C-0666.

\end{document}